# The Identity Crisis
# Security, Privacy and Usability Issues in Identity Management

Gergely Alpár · Jaap-Henk Hoepman · Johanneke Siljee

January 2, 2011

**Abstract** This paper studies the current 'identity crisis' caused by the substantial security, privacy and usability shortcomings encountered in existing systems for identity management. Some of these issues are well known, while others are much less understood. This paper brings them together in a single, comprehensive study and proposes recommendations to resolve or to mitigate the problems. Some of these problems cannot be solved without substantial research and development effort.

## 1 Introduction

Identity management consists of the processes and all underlying technologies for the creation, management, and usage of digital identities. In practice, it covers the process of establishing the identity of a remote user (or system), managing access to services by that user, and maintaining identity profiles concerning that user. As such, identity management is an essential component for the successful development and growth of the next, so-called "2.0", user-centric Internet services. Secure, reliable and user friendly identity management is also considered fundamental in establishing trust, for instance in e-commerce applications [13].

Unfortunately, identity management is also a confusing concept, mainly because the different stakeholders involved (users, service providers, and others) have different views and requirements. This has resulted in quite a number of different approaches towards providing identity management. Several competing systems exist, most of which are in fact under active development. Their features change from time to time, adding to the confusion surrounding identity management.

The historic development of identity management partly explains how this confusion arose. The scope of identity management used to be on a single organisation, managing a limited set of services and employees, specific to one application or ICT platform. Currently this is no longer true. Organisations deliver ICT services to their customers and employees of other organisations as well. This turns identity management into a complex process that has to deal with many applications spanning multiple organisations, instead of one application within one organisation.

The user perspective has also grown in importance. With the increasing presence of organisations on the Internet, and with the creation of a slew of web applications like social networks, web 2.0 mash-ups and the like, users start having their own demands for identity management on the web as well (cf. [22, 20]). For them, managing and remembering a large number of different user accounts on such web sites is cumbersome. Entering name, address and phone number over and over

This research is supported by the research program Sentinels (`www.sentinels.nl`) as project 'Identity Management on Mobile Devices' (10522). Sentinels is being financed by Technology Foundation STW, the Netherlands Organization for Scientific Research (NWO), and the Dutch Ministry of Economic Affairs.

Gergely Alpár
TNO (`gergely.alpar@tno.nl`), and
Institute for Computing and Information Sciences (ICIS), Radboud University Nijmegen (`gergely@cs.ru.nl`).

Jaap-Henk Hoepman
TNO (`jaap-henk.hoepman@tno.nl`), and
Institute for Computing and Information Sciences (ICIS), Radboud University Nijmegen (`jhh@cs.ru.nl`).

Johanneke Siljee
Landcare Research (`siljeej@landcareresearch.co.nz`).

again with every e-commerce site should be avoided. And finally, the identity management systems to support their use of web applications in a variety of contexts should be privacy friendly.

Current systems for identity management do not satisfy these requirements yet. Apart from the fact that properly implementing an identity management system spanning multiple organisations is very complex, the design of current identity management systems is already broken at a more fundamental level. They suffer from several shortcomings that need to be addressed before they can be considered truly secure, privacy friendly and usable. Some of these issues are well known, while others are much less understood. This paper brings them together in a single, comprehensive study and proposes recommendations to resolve or to mitigate them, in order to end the current identity crisis.

1.1 Reading guide

The paper is organised as follows. Sect. 2 gives a short overview of identity management and the state of the art in identity management research. In Sect. 3 we describe fundamental shortcomings in current identity management designs. These concern the concept of identity itself, and the sometimes implicit trust assumptions on which these designs are based. We continue our investigation discussing security (Sect. 4), privacy (Sect. 5) and usability (Sect. 6) issues, and wrap up with an overview of our conclusions and recommendations in Sect. 7.

2 On Identity management

Identity management (or IdM for short), consists of the processes and all underlying technologies for the creation, management, and usage of digital identities. In a typical identity management system we can distinguish three parties: users, identity providers and relying parties[1]. The user (U) requests a service from the Relying Party (RP) that relies on the Identity Provider (IdP) to provide authentic information about the user. These are three technical components, which cannot be held legally accountable. We therefore use the notion of *domain* to represent a legal entity (an organisation or individual person), that is responsible and accountable for the activities of a technical component.

In this paper we loosely define identity as follows. The *identity* of an entity within a scope is the set of all characteristics (also called *attributes*) that have been attributed to this entity within that scope [19]. An *identifier* uniquely identifies an entity (a person, a computer, an organisation, etc.) within a specific scope. We will come back to this distinction later on in Sect. 3.1.

Several types of identity management systems exist [27,30]. We choose to make the distinction between *network-based* identity management and *claim-based* identity management (see Fig. 1), because their difference in architecture has an impact on the security, privacy and usability issues associated with them (cf. [34]).

In a network-based IdM system, the procedure to access a service and to determine the identity and attributes of the visiting user roughly runs as follows. When the user visits the RP, the RP asks the user to authenticate himself at the IdP. The IdP performs this authentication, and if successful gives the user a *token* that the user forwards to the RP. The RP verifies the token, and if valid, accepts the user as authenticated. To obtain further identity information about the user, the RP contacts the IdP directly, using the token as a pointer to the user profile stored by the IdP. In some cases, the user mediates this exchange of information between IdP and RP.

Examples of network-based identity management systems are OpenID[2], the Liberty Alliance[3], and Shibboleth[4].

In claim-based IdM systems a RP specifies the user information it needs in order to grant the user access. The user decides if and how it will comply with that request, by obtaining so-called *claims* from IdPs. A claim is a statement about a user (similar to an attribute assertion in SAML 2.0), expressed (and signed) by an IdP. To obtain such claims, the user needs to authenticate himself to the IdP, and after receiving the claim from the IdP the user forwards the claim to the RP.

The crucial difference with network-based IdM systems is that there is no direct exchange of information between RP and IdP, giving the user more control over the exchange of his identity information. Even though there exist policy tools such as uApprove[5] for network-based IdM systems that allow a user to deny or give consent to releasing his attributes to a RP, the actual attribute assertion exchange still takes place by the RP and IdP communicating with each other directly.

Examples of claim-based identity management systems are the Identity Metasystem (Windows CardSpace) [9,26], and more privacy friendly concepts from the academic community like Idemix [6,5] and U-Prove [4]. In the latter two cases, claims are in fact anonymous, and are not transferred to the RP directly. Instead, the statement in the claim is proven to the RP in a zero-knowledge fashion. This further protects the user's privacy, because it makes the user unlinkable between two interactions with a relying party.

---

[1] Relying parties are also known as service providers (SP).

[2] http://openid.net/developers/specs/
[3] http://www.projectliberty.org/
[4] http://shibboleth.internet2.edu/
[5] Developed for Shibboleth by SWITCH: http://www.switch.ch/aai/support/tools/uApprove.html

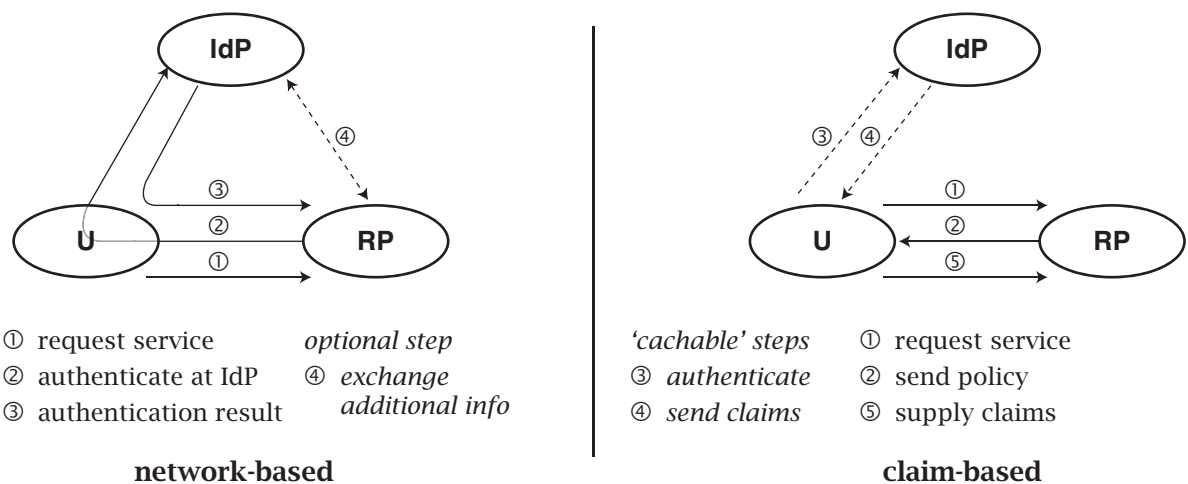

**Fig. 1** Types of identity management systems.

The main points raised in this paper apply to all current types of identity management systems, although to claim-based approaches to a lesser extent. In general claim based approaches exhibit less privacy issues, and have a slightly better security and usability profile.

2.1 Federated identity management

The concept of federated identity management is sometimes cause for confusion. At times the term is used to describe the collaboration of several RPs to use a single IdP, all within the same domain. In our view such a setup is the standard form of identity management, where no real federation takes place. Instead, federated identity management is a setup where identity is shard across domains [25,18]. Within such a federation, additional agreements can be made for further optimisation, e.g. to have a centralised authentication authority. The so-called circle of trust (CoT) equals the set of domains that belong to one federation. Note that a domain can belong to several federations and therefore can belong to several circles of trust. Fig. 2 shows the differences.

Example federations are national education and research federations based on Shibboleth (e.g. Austria's SWITCH, UK Federation, Australian's AAF, USA's InCommon), or Liberty Alliance (Geofederation). Much work is undertaken on inter-federation: technologies and policies to allow users from one federation to be accepted by another federation. This takes place both technologically, e.g. Microsoft's Geneva inter-operates with Shibboleth, and on a policy level, to let e.g. national research federations share resources[6].

2.2 Related work

Several other studies have stressed the importance of privacy, security and usability of identity management, each focusing on specific issues or looking at the problem from a particular perspective.

Pfitzmann and Hansen [33] have been collecting and developing a consolidated terminology about the fundamental concepts in relation to digital identity and identifiability since 2000; the evolving paper is currently at its 34th version.

Public awareness about what private information can be stored and resold by RPs is very low and the customers' view is more optimistic than reality according to the survey by Turow *et al.* [36] Nevertheless, customers do care about their private data and they are willing to take privacy into consideration in purchasing decisions when information about the privacy statement of the retailer is easily accessible and sufficiently user friendly [35].

Paul De Hert [12] argues that a paradigm shift is necessary in connection with private data. Retailers, governments and other organisations have to accept that private information is ultimately owned by the individual who has to be assigned the control over her private data. Privacy has to be a part of the legal framework as well as when designing new systems that comprise personal data. The fundamental technical means, the cryptographic tools, are available to build privacy-enhancing systems with anonymous but accountable users [6,4].

---
[6] see TERENA REFEDs at http://www.terena.org/activities/refeds/

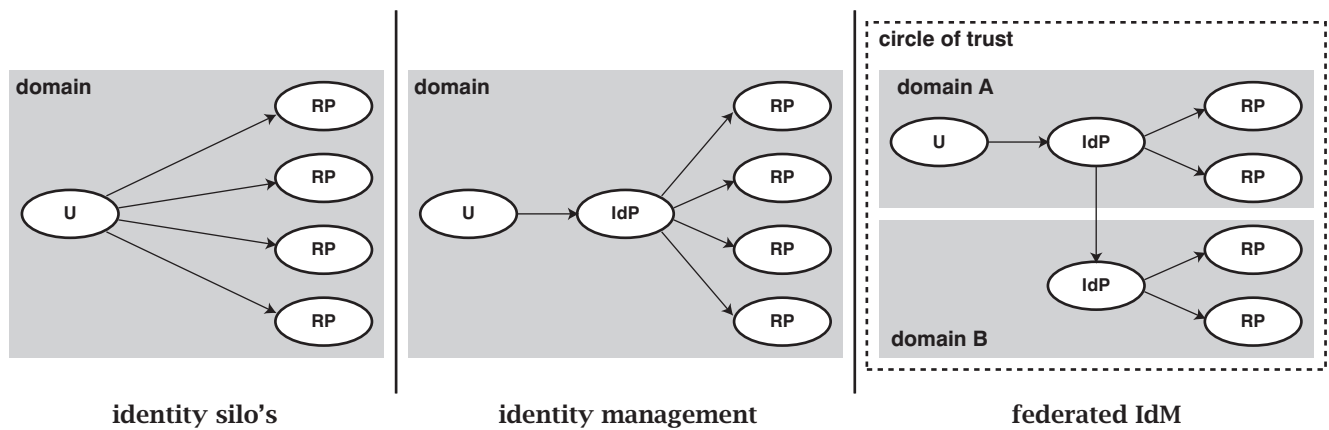

**Fig. 2** Federation in identity management.

Privacy itself also raises many questions to study. Recent endeavours show the different aspects that identity management has to take into consideration. Pearson [31] collects design guidelines for cloud computing services with proper privacy protection and she describes some open questions (e.g. policy enforcement, determination of data processor, constructing privacy design patterns). Another emerging research field is the challenge of building life-long privacy [32] that includes the expansion of solutions to most areas in life and very long-term data security too.

Dhamija and Dusseault [14] provide guidelines about how to design a decentralised web identity management system that will take all participants' motivation as well as their capabilities into account. They assert that the usability aspect is essential in order to achieve wide acceptance and secure usage of such a system, allowing users to take appropriate privacy decisions.

Although federated identity management solutions are widely employed in corporate and academic environments, many problems still arise. These systems can provide convenient user functions (such as single sign-on or automated form-filling), however, the single layer of authentication decreases system security [1] while it increases the value of user credentials (as it provides access to more resources) [14].

One of the most challenging research tasks is how to build a privacy-friendly IMS with good usability properties. Technical research recommends usable privacy-enhancing solutions. Josang *et al.* [20] proposes a scheme that includes a personal authentication device (PAD) that claims to be able to support both secure single sign-on and protection against phishing attacks. A similar tool, a "smart client", is predicted to gain increasing importance that assists the discovery of appropriate IdPs in complex federated systems [25]. In the European PRIME project Camenisch *et al.* [7] developed an elaborate system that provides more control for users about their personal data by automated negotiation processes. Eclipse's Higgins[7] — with a practical open-source approach — is a project in progress that is aiming for implementing a user-centered identity framework for diverse platforms with a consistent user interface.

Ongoing research in IdM encounters many challenges concerning the balance among security, privacy, usability. A suitable legal framework is required that works together with technical solutions [12] and which provides liability incentives [24] for stakeholders. A usable solution for mutual authentication is still to be developed in which not only users are required to provide credentials but IdPs and RPs are also authenticated to the users [20,24].

The Future of Identity in the Information Society (FIDIS)[8] Network of Excellence provides a wealth of information on the topic, see for instance [27] for a systematic review of current systems for identity management. Based on their experiences within that project, Cameron *et al.* [10] propose a framework for a user centric, privacy friendly, IdM, with a focus on ensuring interoperability. Their proposal is very much in line with the US National Strategy for Trusted Identities in Cyberspace [13].

## 3 Fundamental issues

There are several fundamental problems with identity management systems that arise from the illusive nature of the concepts of identity and trust. Also, too lit-tle consideration has been given to the different types of access rights that must be enforced through identity management, as they prove to have an impact on the trust relationships between the parties involved. Because of their fundamental nature, these issues apply to all current models of identity management, and not just the current implementation of such models. We discuss these issues in this section.

---
[7] http://www.eclipse.org/higgins/
[8] http://www.fidis.net

3.1 What is identity?

We first turn to the concept of identity itself (cf. [3]).

Note that identity is *not absolute*. An identity describes an *entity* (a person, a computer, an organisation, etc.) within a specific *scope*. More formally: The identity of an entity within a scope is the set of all characteristics that have been attributed to this entity within that scope (cf. [19]). For example, you may have one identity within the scope of your job, containing information such as your employee number, and another identity within the scope of your family, containing information on the food you like. Identities are therefore only valid within a specific scope. If an identity contains many characteristics, it may uniquely identify a particular entity within a scope. However, with only a few attributes, many entities are likely to match.

It immediately follows that entities have, in general, *multiple* identities. These identities may partly overlap, but can also be mutually inconsistent. One of the authors has blue eyes in all scopes, but may go by different names, nicknames, in different scopes. In extreme cases, people are known to live parallel lives. Sometimes, hardly anybody knows that particular identities in different scopes belong to the same entity.

Identity is *not unique*. Even within a single scope, people may have several different identities. Within the scope of a family a person may not only be a father (to his children) but also a husband (to his wife). Moreover, the identity of an entity is perceived differently by different people, or perceived differently by the same people at different times or in different contexts. Someone may be trusted by one person, but not by another, or only within a certain context.

To uniquely identify entities, one needs to rely on *identifiers*, not on identities. This distinction between identity and identifier is important, and not always properly understood. The confusion is understandable, because in common parlance identity is almost synonymous with personal name, which in turn is understood to be a unique identifier. Note that also identifiers (such as a user name) are only valid and guaranteed to be unique within a scope.

Virtual identities, in the virtual world, can be connected to entities in the real world, but this connection may be loose. For example, computers behind an IP address may be replaced. Ownership of game characters or avatars may be transferred between people over time. In fact, there is quite a large amount of trade in such virtual identities. Likewise, functional roles within companies may look, to external observers, as entities with a particular identity, but different people may actually be assigned to such a role over time.

Identity is also *dynamic*. Assertions about someone's age change when time passes. Your financial situation changes over time, so do your friendships, your convictions and beliefs. Identity management systems must deal with such changes efficiently, and must avoid keeping old invalid data.

Identities may exist long after an entity ceases to exist. The *lifetime* of an identity does not correspond to the lifetime of the associated entity. Most of the time identity information is not updated or deleted after it has become inapplicable. This introduces a privacy risk. But sometimes claims about an entity actually need to be kept long after the entity itself disappears. For accountability reasons, relying parties store usage information for a period of time, sometimes several years. The situation is reminiscent to the difference in lifetimes between keys and certificates (themselves a possible part of an identity). A certificate needs to be kept long after the key it certifies has expired, to allow parties to verify the signatures made with that key.

Identity is not only what you want to reveal about yourself, but also what others conclude, believe, and find out about yourself. In fact most of a person's identity is of this type. Such data may be wrong, become invalid over time, be misrepresented, or be misguiding, etc. In other words, an identity does *not necessarily correspond to reality*. Moreover, it shows that an identity *has many owners*: it is not only owned by the entity it describes, but also collected and owned by others. A fine example of this are your health care records that are being collected by GPs, specialists and other health care personnel. Health records are owned by (and the responsibility of) the GP. You may have the right to *view* them, but you don't necessarily have the right to *change* them. This has important privacy ramifications.

Instead of an entity having one single identity containing all characteristics taken from all scopes, it is more natural to view an entity as a collection of multiple identities (a set of sets), each with their own scope. Note that this aligns with the idea that privacy ensures that information about a person does not leak from one scope into another.

When scopes merge (e.g. if organisations merge) identities may clash. If an entity has an identity in both scopes they may not get merged at all, and as a result the new scope perceives two entities where there is only one. For example, a person may have an account with two different RPs, which require the user to use different IdPs. How to determine what an entity's identity is in the new scope when the two RPs merge? Or when the two IdPs merge?

The fact that identities remain to exist long after the entity 'dies' can result in a wealth of personal information stored in many places, leading to privacy risks for users that are somehow related to this entity. It may also result in IdPs giving out incorrect claims, damaging their reputation of a trusted partner that needs to be right always. Furthermore, claims (that link some identity information to an identifier) may continue to exist indefinitely, even after the identity information itself is deleted. When the claim of an old identity still exists and a new identity is created with the same identifier, these two may seem to refer to the same entity, while this is not the case.

Managing identities does not only mean handling new and fixed identities within one scope, but also handling the complex situations of changing identities in changing scopes, and managing the different perceptions of identity within the same scope. This is a challenge.

*Recommendation* A proper model for identity underlying identity management should be developed, and IdPs and RPs should make explicit how that model applies to their systems of identity management.

Identity management systems should distinguish between the lifetime of an identity, and the lifetime of claims derived from that identity. They should also provide a way to remove obsolete identities (or part of identities) and to invalidate out-of-date claims. Identity management systems should use proper identifiers that satisfy the requirements from Joosten *et al.* [19].

To deal with dynamic identity aspects, it would be convenient if a person could get an attribute certificate for, for example, the date of birth, which could then be used to prove that the person is older than 18 (without revealing the real age). An example implementation of such a system is Idemix [5].

3.2 Different types of access

Identity management systems are being used to enforce different kinds of access rights. These access rights have different risk profiles, and therefore assume different trust relationships between users, identity providers and relying parties. Unfortunately, users as well as system designers are unaware of this difference in access rights. This results in unacceptable risks.

The essential distinction one needs to make is between *membership* and *ownership* of a resource.

Identity management systems were first applied in organisations (to centralise access rights management to business applications) and education (to grant students access to the wireless network, the digital library and the computing facilities). In both cases, the identity management systems are used for deciding whether a certain user is a *member* of a group. In the first case it decides whether the user is a member of the group that has access to some business application. In the second case it decides whether the user is a student of a certain university. The resource being controlled is not owned by the user, and any risks or resource damage due to using the identity management system lies completely with the relying party, not the user.

More and more, identity management systems are also being used to enforce *ownership* of a resource. The prime example is on-line banking, and to a lesser extent email, chat, blog and social networking accounts. Illegal access to your bank account will hit you with a direct financial loss. Access to your email, chat and other systems may enable a criminal to 'steal' your identity, which may hurt you in many other ways. In this case, the risk of using the identity management systems lies completely with the user.

How does this affect the use of identity management systems? To enforce membership, identity management needs to assume different trust relationships than to enforce ownership. In the first case, the relying party needs to trust the identity provider to reliably authenticate its members. In the second case, it is the user that needs to trust the identity provider to reliable authenticate himself. These trust relationships need to be enforced either by technological means, or through mutual agreements like service level agreements (SLA) with associated penalties. In either case, an identity management system to enforce membership is inherently different from an identity management system to enforce ownership.

Also the risk level associated with using identity management differs. In the case of granting students access to university resources, the damage associated with abuse (and therefore the risk of using identity management systems) is quite low. Except for extreme, denial-of-service cases the university does not suffer any direct actual loss of non-students having access to the resources. This is the same for any *subscription* based digital service, like on-line music, or a digital newspa-per, etc. Because the marginal cost of the copy is essentially zero, there is no direct loss if non-members have access as well. The losses incurred by such services are the result of fewer sales.

Granting access to business applications (and the associated data in particular) has a higher risk profile. Not because of loss of revenue, but because the data is often confidential. It could cause enormous financial damage when it becomes public. Similarly, there is a difference in risk level associated with granting access to a bank account and granting access to an email account.

*Recommendation* The impact of dealing with different types of resources on IdM deserves further study. For instance, related distinctions one could make here are on *rivalry* and *durability* of a resource. A *rivalrous* resource cannot be used at the same time by another user, whereas access to a *nonrival* resource does not exclude such access by others (cf. common property resources). *Durable* resources do not degrade or get used up, whereas *non-durable* do degrade or can be used only once (cf. the difference between 'bits' and 'atoms'). It is interesting to explore the economic literature to see whether even more types of resources and goods can be discerned, and how they influence the trust assumptions in (and the risk of using) identity management.

3.3 Trust assumptions

We have been using the ambiguous concept of trust in previous sections, without giving a definition. We will not present a thorough discussion of the notion of trust in this paper though, but refer to Hardin [17] O'Hara [29] and others[9]. For our exposition the following informal definition (taken from van den Broek and Huijboom [37]) is sufficient:

> When an actor trusts another actor, he or she is willing to assume an open and vulnerable position. He or she expects the other to refrain from opportunistic behaviour even if there is the possibility to show this behaviour.

In more technical terms, entity A trusts entity B if B can break the security or privacy policy of A without A's cooperation or knowledge. Similar definitions can be found elsewhere (cf. [21]).

3.3.1 Building trust

Trust can only be built over time. For this, the RP needs to be sure it is talking to the same entity (and the other way around) in different sessions. In order to do so, both parties need to retain information from session to session. Unfortunately, in many of the current IdM systems, the User does not maintain any state. Moreover, the RP is completely relying on the IdP to ensure that the link between different visits of the same user is reliable[10] (but see also [16]).

The "proof key" of CardSpace [26] does not solve this: this only prevents an adversary to use a security token it obtained illegally[11]. The binding is only guaranteed as long as the IdP is honest. If the IdP releases the private proof key, or if it uses that proof key itself, the UA is no longer involved.

The problem could be solved if the User Agent and the Relying Party each store part of a key pair, and verify the link directly without external help.

3.3.2 Trust assumptions are ill understood

By using an identity management system, one implicitly agrees to several complex, and poorly understood, trust relationships between the parties that belong to that identity management system. Some of the trust relationships involved in identity management are the side effect of more fundamental security and federation problems, that we will discuss separately.

*The user trusts the IdP not to act on its behalf without his explicit consent.* In many systems for identity management, the IdP essentially does the logging in to a RP, on behalf of a user. It could easily do so, without the user even being present. Clearly the user does not want the IdP to do this. The impact of this concern is unclear (an IdP that betrays the trust of user is soon out of business), but a fix to prevent this scenario is not difficult to implement (see section 4.1). Additionally, the user expects the IdP not to release personal information unless explicitly asked by RP and with the permission of the user.

---

[9] Lacohee, H. Crane, S. and Phippen, A. Trustguide: Final report, at http://www.trustguide.org.uk.

[10] This is also a problem with current PKIs, where the RP also does not keep state and trusts the certificate coming with an authentication to ensure a long term binding between several encounters with the same user over a long period of time.

[11] It works as follows: the proof key pair is generated by the IdP. The IdP sends the private proof key to the User Agent (UA), encrypted using the public key of the UA. The public part is also sent to the User Agent, together with the security token. The entire message is signed by the IdP. Using the private key it received earlier, the UA generates a signature over this combination of token and public proof key, and sends that to the RP. The signature proves to the RP that the UA knows the corresponding private proof key.

*The relying party trusts the IdP not to extend the circle of trust (without his consent).* Depending on the application, a RP may rely on an IdP to provide him with attributes regarding the user accessing the service. Based on these attributes the RP may decide to grant the user access or not. A common example is granting access to a wireless campus network to all students, including those that come from other universities. In this application, the IdP will tell the network whether the user is a student or not. The circle of trust could be extended when a new university wants to join the scheme. In this case the IdP will delegate the responsibility to classify the new members as students to the newly connected IdP, and forward this classification as its own to all RPs that connect to it. Based on the decision of this *new*, and unknown, IdP the network (by necessity) will grant these users access to its network. The decision to grant access to a user is thus in the hands of the new IdP, which may be undesirable. See also section 4.4.

But there are many other trust assumptions involved in IdM. The most basic trust relationship underlying identity management (and this is usually well understood) is the following.

– The relying party trusts the IdP to make a particular claim about a particular user.

However, the following trust relationship is equally important and fundamental, yet most people will not realise this assumption is being made.

– The user allows the IdP to make a particular claim about herself to a particular relying party, and allows the relying party to accept such claims from this IdP.

Note also, that these trust relationship are dynamic and context dependent: a user may at some point decide to no longer use the services of an identity provider, and therefore the trust relationship no longer exists. Moreover, the user may only allow the relying party to accept certain claims from the identity provider within a certain context. For example, if a user only accesses a service from work, or during the day, the relying party should not accept claims about the user during the night, or when it appears the service is accessed from an Internet kiosk.

Every trust assumption is a potential security problem, as the trusted party can break the security policy of the other party. From a security point of view, it is preferable to rely on as few trust assumptions as possible.

*Recommendation* A better understanding of the trust assumptions among the parties involved in an identity management system is needed. More implicit or explicit trust assumptions should be collected and studied, and it should be determined whether they can by mitigated or avoided by other (e.g. technical) means.

## 4 Security issues

Current identity management frameworks have implemented techniques, methods, and policies to securely handle identity information. However, several vulnerabilities remain.

### 4.1 The IdP is a single point of failure.

Identity management systems require the user *and* the relying party to place a large amount of trust in the IdP (see also Sect. 3.3). A wealth of identity information is stored at IdPs, and users can do nothing but simply trust the IdP to preserve their privacy and properly secure their identity information [14]. But still, mistakes can be made and privacy-sensitive information can become public[12]. This makes the identity provider a single point of failure.

Possibly even more worrisome is the fact that in most current identity management systems the IdP has all information it needs to log in at related RPs as a registered user. This means that anyone that has access to this information at the IdP can log in as a user at the related RPs: For example employees of the company hosting the IdP, or hackers that break into the IdP systems. Depending on the service, the impostor could order things to make the legitimate user pay money, transfer money from the user's bank account, or get insight into personal information, such as the user's electronic health record. The RP has no means to distinguish the impostor from the real user.

This feature can also be (ab)used to turn an IdM system into a system for mass surveillance. If the identity provider happens to be the government (and many governments offer IdP services and actively try to extend their use in other domains), then the government has immediate access to all your data stored at services that accept this IdP. Using such an IdP to manage your identity at your bank, your ISP, or other relying parties is not recommended in such systems.

The possibly large collection of data stored at an IdP can also be used to perpetrate identity fraud. If information about a user stored at the IdP becomes public due to e.g. theft, hacking or implementation flaws, this data can be used to fake an identity when registering for a new service.

---

[12] Privacy Rights Clearinghouse, "Chronology of Data Breaches" http://www.privacyrights.org/data-breach.

The impact of this issue increases as more and more systems get federated and a single IdP is used to access a large number of services. Such an IdP may abuse its powers, maliciously accessing many services, before a user notices. If the RP and the IdP do not properly log authentication requests and access control decisions, both the RP and the IdP may claim that the other party was at fault, and the user will not have any evidence to determine what actually happened.

*Recommendation* To prevent this issue it is necessary to put the user in control of information that is released from the IdP [8], not only by policy, but technically enforced into the identity management system[13]. It should not be possible for the IdP to log in to a RP claiming to be another user. The identity management system should enforce the requirement that the user controls part of the data necessary to log in to the RP. Cardspace [11] tries to achieve this using the concept of a Private Personal Identifier (PPID) [2], and the use of a proof key (cf. Sect. 3.3.1). This does not solve the problem completely, because the server needs to remember the associated public keys in order to detect spoofing attempts using the same PPID but with freshly generated key pairs. Also, this approach breaks the 8th law of identity (see section 6.1, and [8]) on location independence as a special user agent needs to be installed on the user's PC.

### 4.2 The risk of phishing is increased

Most current identity management systems only provide a way to authenticate the user, but it is not possible for the user to authenticate the IdP or the RP [14]. This is a necessity to be able to prevent phishing attacks, where attackers trick users into revealing identity data and credentials. With identity management becoming more wide-spread, phishing attacks based on getting IdP login credentials will most likely increase as well.

When HTTP redirects are used (as for example in OpenID 1.0) phishing attacks are even easier to launch[14]. It is as simple as creating an illegitimate, but attractive, website that redirects to a false copy of the IdP to capture the user's credentials.

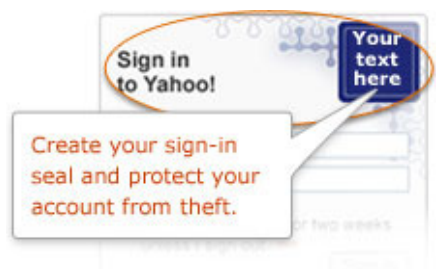

**Fig. 3** Yahoo! sign-in seal to protect against phishing.

An example countermeasure to phishing attacks using fake IdP websites is Yahoo! sign-in seal[15] (see Figure 3). This is a personalised image or short text phrase that will appear each time a user logs in to a Yahoo! web page from the same computer the seal was created on. The presence of the seal enables the user to distinguish the real Yahoo! sign-in page from a false page. This solution only works if the user logs in using the same computer as the seal was created on, as Yahoo! identifies it by storing tags in multiple places on the computer[16].

*Recommendation* To prevent phishing attacks it is very important that users can (and will) authenticate the RP and the IdP. Mutual authentication therefore needs to be incorporated in identity management systems, in such a way that the user is not required to install special software or to use one and the same computer all the time (as is the case with Yahoo! sign-in seal and Microsoft Cardspace). Furthermore, authentication of the IdP and RP by the user should be more user friendly than checking their SSL certificate manually. There does not appear to be a single, usable, and secure fix to prevent phishing in all cases.

### 4.3 What is the optimal size of a key chain? – *or* – How many identities should a user have?

One of the main advantages of identity management for end-users is single-sign on: not having to remember all those user names and passwords, except for the log-in-token for the IdP. From this perspective, it would be great to have just one IdP: only one user name/password (or another authentication token) and that's it.

Of course, this is not feasible. Not only because users may not trust that one IdP to have access to all their services (see section 4.1). Even if users *do* trust a single

---

[13] Other systems allow for users to give their approval and store that approval as a policy setting at the IdP. An example is uApprove, an optional extension for Shibboleth. uApprove is not location dependent, but has its own limitations: it is optional, and the control over that tool and its data (e.g. stored user consents) still resides at the IdP.

[14] M. Slot, "Beginner's guide to OpenID phishing", http://marcoslot.net/apps/openid/, retrieved in December 2008.

[15] https://protect.login.yahoo.com/.

[16] "How does Yahoo Sign In Seal Work?", blog, retrieved December 2008, http://girishnayak.blogspot.com/2006/09/how-does-yahoo-sign-in-seal-work.html.

IdP for that, using only one IdP means that if that IdP is compromised, all identity data is compromised immediately as well. It is therefore advisable for users to distribute their identity information over multiple IdPs. Furthermore, different RPs will require different IdPs. Financial institutions for example have other requirements and preferences than car rental agencies with respect to an IdP. The first may want to set up their own IdP to be able to control the security of authentication, while the latter is satisfied with using a third-party IdP. Can we then settle for one IdP for personal use, one for work, and one for each financial institution? This seems to be a workable yet quite arbitrary subdivision.

The question is: how many identity providers does a user need? What is the best compartmentalisation of the digital identity mess? We need to understand the advantages and risks of using a certain amount and distribution of IdPs and federations, in terms of security, usability, and business.

*Recommendation* To be able to determine which and how many "identities" are optimal, a model that captures these relevant aspects needs to be developed. To our knowledge such a model does not exist yet.

### 4.4 Federations are risky

In cross domain settings, one organisation may assign roles to certain individuals, while another organisation assigns access rights to roles. This is typically done in federation settings: one university classifies certain people as students of that university, while other RPs rely on that classification to mediate access to resources like the library, classes, or the student restaurant.

This gives rise to a compliance defect [15]: the IdP may interpret the semantics of the role (e.g. when someone classifies as a student) differently from the RP, which leads to a situation where a person gets access to a service that he or she is not supposed to access. The reverse (being denied access) is a problem as well.

The above is an instance of a more general issue. Traditionally, access to a resource or service is mediated through a "reference monitor" [23]. In an identity management system, this reference monitor is in a sense distributed over several parties. The underlying question is how to do this "split". In the simplest case, this question surfaces as the question "where to keep the access rights".

A separate issue is the control over the so-called Circle-of-Trust (CoT). By establishing a federation among several IdPs, the CoT is similarly extended. RPs connected to a certain IdP may have limited control over this, and therefore have limited control over the risks that they are exposed to because of the extension of the CoT.

*Recommendation* When implementing or joining an identity federation, RPs need to carefully consider where to keep and maintain the access rights. Moreover, they need to judge the consequences when the CoT is extended without their knowledge or consent.

## 5 Privacy issues

Identity management systems are used to facilitate millions of user transactions on the net each day. They mediate between user and relying party, handle a lot of personal information, and often register who does business with whom. This has obvious privacy consequences. We discuss these issues in detail below.

### 5.1 Linkability across domains

Like AdSense[17] and DoubleClick[18], identity management systems have the potential to track a single user over all the websites he or she visits.

To maintain privacy, it should be possible for users to be anonymous or use pseudonyms at RPs, and to choose IdPs that do not link all user transactions at all RPs together, and so do not keep records of everything each user has been doing. Many identity management systems implement at least part of these solutions, which is why the UK Information Commissioner has recognised Federated Access Management as a Privacy Enhancing Technology[19]

However, not all identity management systems do. An example is DigiD, the Dutch national authentication provider that enables authentication of Dutch citizens when communicating with Dutch government institutions. DigiD uses the BSN (Burger Service Number) to identify a user: after authentication DigiD sends the BSN to the RP. This number uniquely identifies each user, and does not allow for anonymity or pseudonymity at all. As indicated in [28], expanding the scope of DigiD to incorporate not only governmental organisations but also the private sector has many advantages. However, as already mentioned, in its current form DigiD does not allow for pseudonymity as the BSN is always used as identifier. Such extension of use of DigiD for the private sector is not acceptable as it violates the user's privacy when all RPs always receive the BSN as user identifier, thereby always knowing exactly who the user is.

---

[17] https://www.google.com/adsense/
[18] http://www.doubleclick.com
[19] See http://www.ico.gov.uk/upload/documents/library/data\_protection/detailed\_specialist\_guides/privacy\_enhancing\_technologies\_v2.pdf

Another example stems from the need to retain user permissions at RPs when a user moves from one home organisation (and thus IdP) to another. Federations often solve this by implementing one static user attribute (often a pseudonym identifier) that a user can 'bring' to another IdP. This 'feature' severely limits the privacy of users, as the static attribute links all user actions at all its previous and current IdPs, and one (but often many RPs).

This also involves a paradigm shift in the applications relying on identity management. It has become standard practice to require a user to identify herself before granting access to a service. In many cases this is unnecessary. For example, in order to be allowed to buy alcohol, someone only need to prove that he or she is over 16 years old (or 18 or 21, depending on local laws). Such 'attribute' or 'credential' based forms of privacy friendly identity management do exist in theory but are rarely applied in practice [5].

*Recommendation* Whenever using identity management systems, one should always try to implement maximum anonymity and pseudonymity where possible. A solution for expanding DigiD to the private sector is to use pseudonyms that are based on DigiD as identifiers. Each user will have a different pseudonym with each RP, and no pseudonym should leak any information about the underlying BSN. A possible method for generating pseudonyms is making a hash of the BSN. Alternatively, if it needs to be possible to trace the pseudonym back to the original BSN, different encryption methods can be used [28]. Furthermore, a privacy friendly method for retaining access rights at RPs when changing IdP is necessary.

5.2 IdP knows all user transactions

In current identity management systems the IdP is involved each time a user authenticates at a RP. Therefore the IdP can keep track of all these user actions (although sometimes the specific RP involved may be kept hidden from the IdP). In most systems the user is not even involved in the exchange of his identity information between IdP and RP. But even in a claim-based identity management system such as Cardspace, where the user needs to give consent before identity information is transferred, even though the IdP does not need to know exactly who the RP is, the IdP often needs to generate the assertion on-line and therefore knows of all user transactions.

It seems PKI is the solution to this issue. Here a Certificate Authority (CA) identifies and authenticates a user only once, and then certifies the user's public key. The user can then authenticate himself to a RP by signing data with his private key, which the RP can verify using the corresponding public key. In this case the CA is not directly involved in the user authentication by the RP, but is still the trusted third party. The main downside of this solution is the necessity for the user to always have his private key certificate available when logging in, and thus PKI identity management violates the 8th Law of Identity (see section 6.1). Also Identity Selectors, as used in Cardspace, violate the 8th Law of Independence, as all identity selection solutions are hardware-specific, OS-specific or even browser-specific.

*Recommendation* We need to develop an identity management system that does not require IdPs to see all user transactions, without violating the 8th Law of Identity. This apparent paradox may be solvable by relying on personal hardware (like tokens, smart cards), or be developing mobile identity management concepts.

5.3 Proportionality and subsidiarity often violated

In the EU, most of the data protection or privacy laws are based on the principle of proportionality and subsidiarity. Proportionality stipulates that the amount of personal data being collected is proportional to the goal for which it is being collected. Subsidiarity demands that the same goal cannot be achieved in a more privacy friendly way.

Often, websites and services violate these principles. You do not need to know someones identity to determine his age. Subscriptions for a service can certainly be handled anonymously. An on-line newspaper does not need to know *who* accesses the system, all it needs to know is whether that person is *entitled* to read the news on-line.

*Recommendation* Less is more. RPs should be precise about the personal information required to offer a service, and should not ask for more information "just because they can". Think about anonymous ways to offer the same service.

**6 Usability issues**

Although many identity management systems claim to be designed with the user in mind, most still have important usability issues. We discuss those issues in this section.

6.1 The 8th Law of Identity: Location Independence

The seven laws of identity [8] present a compelling set of requirements a system for identity management must satisfy. However, one important usability aspect is missing, which we pose as the 8th law of identity:

**Definition 1 (Location Independence)** The identity system must allow a user to create, manage, and use his identity independently of his current location and current device in use.

A user should be able to access a RP using the identity management system not only from his PC, but also from a computer at a cybercafé in Hong Kong.

*Recommendation* Identity management systems should not rely on any persistent data stored locally at the user's machine. Note that this recommendation contradicts some of the other recommendations in this paper. Exceptions to this rule could be hardware tokens that can easily be carried with you to achieve higher levels of authentication.

6.2 "Who am I today?"

As discussed in section 3.1, users may have several identities, even within a single scope. This distinction in identities manifests itself when people have several different responsibilities, or, in other words, may have several different "roles". Examples may help to clarify this issue.

When signing a document, a notary can choose to sign this as a notary, or as a private person. The distinction is legally significant. The CFO of a company may use an electronic banking system either to enter a personal or a business transaction. An ICT system administrator may sign in to a system either as "root" (which allows him to run OS-level applications and scripts) or as an ordinary system user (that allows him to only execute end-user applications).

We see that users can have different roles that allow them to do different things within a certain service. Furthermore, the impact of user actions depends on their role: a signature of an accountant or a notary represents more legal value.

Current identity management systems do not make it easy for users to manage such different roles (although, to be fair, exceptions exist [11]). Basically, users are forced to maintain and manage several identifiers to separate these roles. But this may lead to confusion. For instance, if a user has previously signed in at its IdP using a particular identity, and the user and the service support single sign-on, the user may automatically be signed in using this same identity when accessing a different service some time later. This is potentially dangerous: if the CFO signed in as CFO earlier, he may not want to execute a personal transaction while still being signed in as CFO.

Depending on the type of service, actions performed in a certain role may be visible to others that can also access that role, or result in information sent to the organisation that is responsible for that role. For example, all communication of the president of the USA is kept for later reference. The same goes for many transactions performed when doing business. In these situations privacy sensitive information can become public, such as the purchase of personal books, visiting certain websites, or the rental of a hotel room, when the role that was selected to execute those actions happened to be a business role.

For many current identity management systems these very common usage scenarios pose a problem. There is no way to indicate as which role, or which identity, a user wants to access a particular service, especially if he has accessed that system in both capacities before. One of those identities may be selected automatically (in a single-sign-on context), most likely without the user knowing why, or how to change it.

*Recommendation* Identity management systems should provide a way for users to see and select their identity with which they "sign in" even if explicitly signing in is not asked for, because the user has already authenticated with an IdP that is recognised by the RP. Asking users each time which role (at which IdP) they want to use is cumbersome for the user, and therefore not a good solution to this issue. So alternative approaches need to be investigated.

6.3 When complex transactions require multiple credentials

A special case of the previous issue is that of transactions that require the cooperation of many services, possibly of multiple RPs. This is for example the case in Service Oriented Architectures (SOAs), where one application consists of multiple software services. The problem arises if the user needs to present credentials for more than one service, and the credentials depend on the role the user assumes. The user needs to have all the credentials required to perform the transaction, but can only present them if he logged in using the right role. Also in this situation the user has no means to select his role or identity for a particular session.

*Recommendation* Clearly this is part of a more general problem of implementing chains of transactions, in which identity management plays a role only partially. But perhaps identity management systems could provide a way to automatically determine the full set of required credentials, and the minimal role the user can assume that covers those credentials.

6.4 User profile management

When a user accesses a service, this often involves the processing of personal information. Some of that information may be stored at the IdP, while other information is stored at the RP, as it is service specific and the RP needs the information for e.g. marketing research, or because the RP does not trust the IdP to store the required information instead. More often than not, many RPs store the same information for a particular user. How can such a scattered profile be managed and be kept up to date by a user? This could be useful for instance to allow a user to update his address every time he moves. Should a user always be allowed to update such information (consider for example medical records)? The question is whether identity management systems will be able to simplify user profile management both for the end users as well as the RPs and IdPs.

*Recommendation* This issue is resolved by following the emerging trend towards the convergence of profile, identity, and authorisation (or access) management into a single system for identity management. Such a system would also be beneficiary for users, as it allows them to manage personalisation of many different services in a central location. This way, changing a single setting once will change the behaviour of all services consistently. This enables a ubiquitous personal experience across many different services.

Note however that although identity management is positioned as solution for the cumbersome maintenance of identity information, the nature of certain business (see Sect. 4.4) and the nature of identity itself (as discussed in Sect. 3.1) limits the implementation of management of identity information across organisational domains.

7 Conclusions & Recommendations

Identity management not only comprises identification and authentication, but also access management and user profile management. Stakeholders such as end-users and relying parties require identity management systems to be able to span multiple organisations, to be user-friendly, privacy friendly, and secure. Current systems for identity management are not able to accomplish this.

As we have seen in this paper, security and privacy are not adequately addressed in current identity management solutions. This renders current systems for federated identity management inapplicable for 'high-value' services like electronic banking, that consequently remain to rely on their own home-grown systems for access control.

Federation, as well as the more fundamental concept of identity, and its consequences regarding scope, responsibility and trust, is still not understood. More fundamentally, the term federation is used confusingly within the field. This should be avoided.

The issues of identity management systems presented in the paper cause the current identity crisis. In order to resolve the identity crisis, we recommend to follow up on the following main observations made in this paper.

- A proper model for identity underlying identity management should be developed, and IdPs and RPs should make explicit how that model applies to their systems of identity management.
- Building on that model, the trust relationships between the parties using an identity management should be investigated and formalised.
- To prevent phishing attacks it is very important that users can (and will) authenticate the RP and the IdP. Mutual authentication therefore needs to be incorporated in identity management systems, in such a way that the user is not required to install special software or to use one and the same computer all the time
- To enhance user privacy we recommend that users can remain anonymous or use pseudonyms at RPs, and to have IdPs that do not link all user transactions at all RPs together. Although identity management systems already implement at least part of these solutions, not all do so. We need an identity management system that does not allow IdPs to see all user transactions, without violating the 8th Law of Location Independence (which states that identity management systems should not rely on any persistent data stored locally at the user's machine).
- Identity management systems should provide a way for users to see and select their identity with which they "sign in" even if explicitly signing in is not asked for.
- Identity management systems should provide a way to automatically determine the full set of required credentials for a certain service, and the minimal role the user can assume that covers those credentials.

- Finally, we need identity management systems that put the user back into control and that support the user in maintaining a user profile that can be used (in a controlled manner) by business from several organisational domains.

Most of these recommendations are not trivial, and to implement them requires a substantial research, development, and standardisation effort. Moreover, to resolve the identity crisis stakeholders need to work together on this. We believe the growing need for a proper, well-founded, identity management solution legitimates the effort.